\documentclass[aps,pra,twocolumn,showpacs]{revtex4-1}
\usepackage{graphicx}
\usepackage{amssymb,amsfonts,amsmath}
\usepackage[colorlinks=true,citecolor=Cerulean,linkcolor=RubineRed,urlcolor=Cerulean]{hyperref}
\hypersetup{breaklinks=true}
\usepackage{graphicx}
\usepackage{color}
\usepackage[usenames,dvipsnames]{xcolor}
\usepackage{epstopdf}
\usepackage{braket}
\usepackage{physics}
\usepackage{dsfont}

\def\to{\rightarrow}
\def\sZ{\mathcal{Z}}
\def\sH{\mathcal{H}}
\def\sX{\mathcal{X}}
\def\sT{\mathcal{T}}
\def\sI{\mathcal{I}}
\def\sD{\mathcal{D}}

\begin{document}
\setlength{\parskip}{1em}

\title{Topology and Interactions in the Photonic Creutz and Creutz-Hubbard Ladders} %Rev1
       
\author{J.~Zurita}
\affiliation{Departamento de F\'isica de Materiales, Universidad
Complutense de Madrid, Pza. Ciencias 1, E-28040 Madrid, Spain}

\author{C.E.~Creffield}
\affiliation{Departamento de F\'isica de Materiales, Universidad
Complutense de Madrid, Pza. Ciencias 1, E-28040 Madrid, Spain}

\author{G.~Platero}
\affiliation{Instituto de Ciencia de Materiales de Madrid, CSIC, C/ Sor Juana In\'{e}s de la Cruz 3, E-28049 Madrid, Spain}

\begin{abstract}

The latest advances in the field of photonics have enabled the simulation of an increasing number of quantum models in photonic systems, turning them into an important tool for realizing exotic quantum phenomena. In this paper we suggest different ways in which these systems can be used to study the interplay between flat band dynamics, topology and interactions in a well-known quasi-1D topological insulator: the Creutz ladder. Firstly, a simple experimental protocol is proposed to observe the Aharonov-Bohm localization in the noninteracting system, and the different experimental setups that might be used for this are reviewed. We then consider the inclusion of a repulsive Hubbard-type interaction term, which can give rise to repulsively bound pairs termed doublons. The dynamics of these quasiparticles are studied for different points of the phase diagram, including a regime in which pairs are localized and particles are free to move. Finally, a scheme for the photonic implementation of a two-particle bosonic Creutz-Hubbard model is presented.

\end{abstract}

\maketitle

%%%%%%%%%%%%%%%%%%%%%%%%%%%%%%%%%%%%%%%%%%%%%%%%%%%%%%%%
%MAIN PART
%%%%%%%%%%%%%%%%%%%%%%%%%%%%%%%%%%%%%%%%%%%%%%%%%%%%%%%%

%SEC INTRO
\section{Introduction}
\label{sec:Intro}

Recent years have seen vast advances in the field of quantum simulation, 
an approach first proposed by Feynman, \cite{feynman1982} in which the 
dynamics of some complicated quantum system is simulated by another,
highly-controllable, quantum system. 
Analog, purpose-built, non-universal quantum simulators using photonic systems and cold atoms in optical lattices have allowed experimentalists to physically realize a wealth of exotic quantum phases and phenomena that were previously only theoretical proposals. In principle, this technique can provide an enormous speed advantage over the conventional approach of simulating such systems using digital (classical) computers, a feature known as ``quantum supremacy''. \cite{preskill} Its analog nature eliminates the need for error correction protocols or high-fidelity quantum gates, making their implementation easier than that of digital quantum computers. \cite{Tangpanitanon2019} It's been estimated that a few tens of controllable qubits would suffice for achieving a significant quantum advantage over a classical computer. \cite{Georgescu2014}

A particularly important application of quantum simulation is the investigation
of topological insulators; materials which are insulating in the bulk, but
possess conductive surface states protected by topological order. \cite{topology}
The simplest examples of this form of material occur in one-dimensional, \cite{SSH}
or quasi-one-dimensional systems. A notable example of the latter type of system is the Creutz ladder, \cite{Creutz1999}
in which particles (bosonic or fermionic) move in a ladder-like lattice in the presence of a magnetic field, as shown in Figure \ref{fig:LattPh}. 
For some special values of this magnetic field, and in the absence of the rung-like links, the energy bands of the Creutz ladder become flat. This means that the group velocity of the particles, $v_g=\partial_k \omega(k)$, vanishes, and that free propagation of single particles is thus impossible. The mechanism behind this localization is the Aharonov-Bohm (AB) effect, the phase acquired by the wavefunction of a charged particle by moving through a region with a non-zero magnetic vector potential. This effect is also seen in some bipartite lattices such as the $\sT_3$ lattice \cite{Vidal1998a} and the rhombus chain, \cite{Vidal2000} and it has been termed ``Aharonov-Bohm caging''. This localization effect has been observed experimentally in superconducting wire networks, \cite{Abilio1999} metal wire lattices, \cite{Naud2001} ultracold atom lattices \cite{Kenji2002} and, most recently, in a photonic system. \cite{Mukherjee2018}

As the single-particle kinetic energy is
zero in a flat band, even extremely weak interactions can have an enormous
effect on the system's properties. Introducing a contact interaction
of Hubbard-type lifts the degeneracy of the flat bands and breaks AB-caging, significantly complicating the system's properties and producing
a rich filling-factor-dependent phase space which can include a quantum pair liquid, a Wigner solid, a condensate, and a supersolid phase. \cite{Takayoshi2013,Tovmasyan2013}
For strong interactions, particles can form repulsively-bound pairs called
``doublons'' which can themselves experience AB-caging at a different value of the magnetic flux. \cite{Creffield2010}
This rich phenomenology makes the Creutz ladder an excellent arena to study the effects and interplay of topology, AB localization and interactions.

Photonic quantum simulators use the properties of quantum or classical states of light to analyze the behavior of different quantum systems. In the last few years, an increasing collection of topological materials have been studied in photonic setups, \cite{TopoPhotoRev} both using quantum \cite{Leykam2015,Mittal2018,Wang2019} and classical light. \cite{Longhi2013,Slobozhanyuk2016,Lee2018,Liu2019} The latter type of simulators are easier to implement but can only simulate first-quantized dynamics, and thus the realization of systems of interacting particles has been traditionally challenging. Nevertheless, interesting advances have been made in the field of nonlinear classical photonics in general, and in strongly correlated systems in particular. Especially relevant for our work are the nonlinear photonic setups which also include topological \cite{Hadad2018,Serra2019} or AB localization effects. \cite{Liberto2018} Additionally, a quantum topolectrical system with interactions has been recently realized experimentally. \cite{Olekhno2019}

For our current purpose, we will focus on photonic waveguide lattices along which classical light propagates in the paraxial regime, where the direction along the waveguide takes the role of the time coordinate in the quantum system. Recently, we have seen the first steps in the use of photonic waveguide lattice simulators for the study of both strongly correlated systems \cite{Longhi2011,Krimer2011,Corrielli2013,Rai2015,Mukherjee2016} and topological insulators \cite{Zhong2017,Noh2018,Zhong2019}. We will propose a setup to include both interacting and topological effects.

In this paper, we first study the effects of AB interference on one-particle states in the Creutz ladder, and show how they can be observed experimentally in a photonic waveguide lattice. We then proceed to investigate how interactions modify these results,
by introducing the minimal model which incorporates interparticle interactions: the two-particle Creutz-Hubbard ladder. Finally, we propose a way to implement the latter model in a photonic waveguide system, focusing on the geometry of the lattice and the necessary use of synthetic dimensions. \cite{Lustig2019}

%SEC CREUTZ
\section{Creutz ladder}
\label{sec:Obj}

\begin{figure}[!htbp] 
    \centering
    \includegraphics[width=\linewidth,trim= 13 10 15 15,clip]{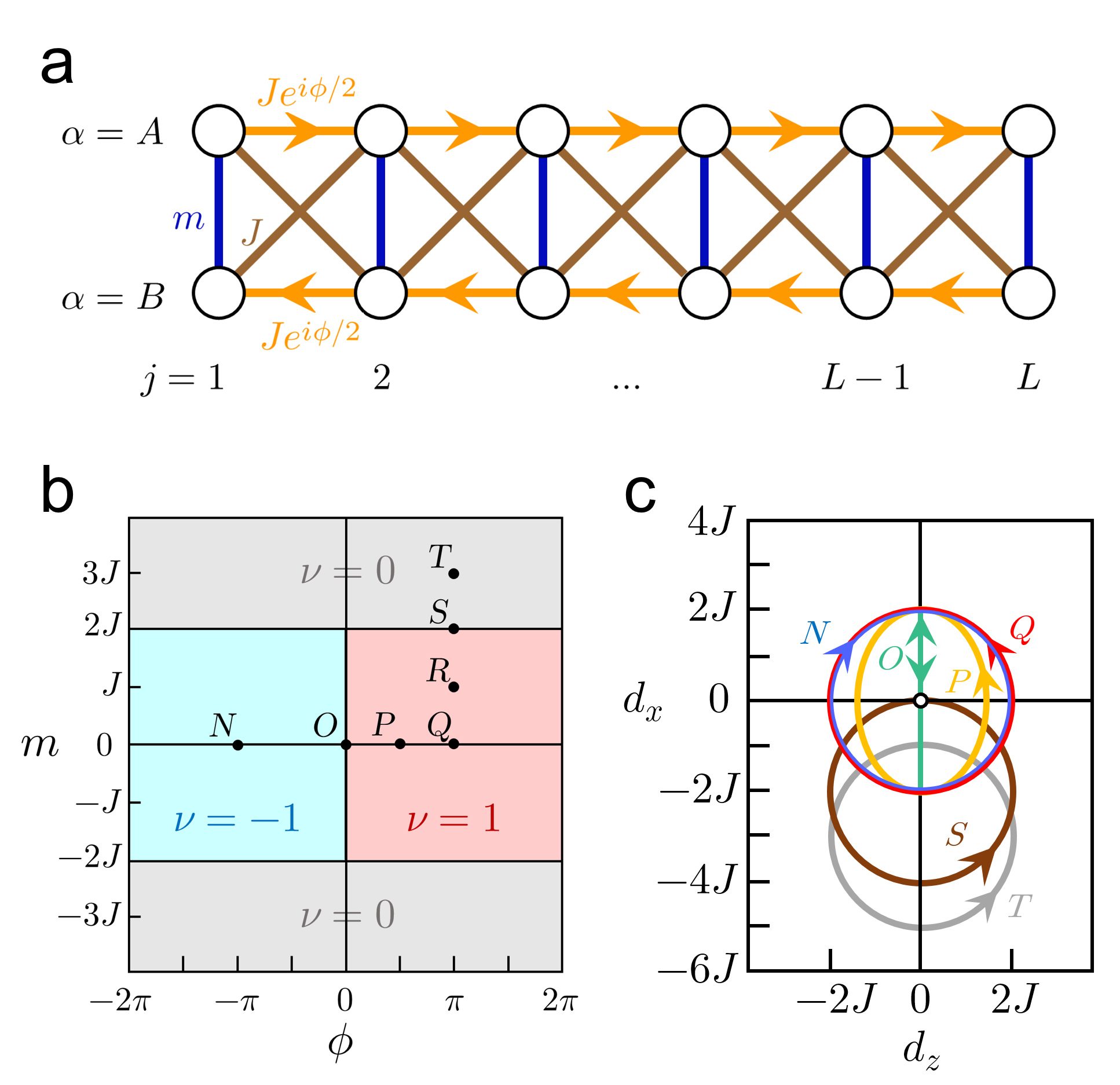}
    \caption{a) Geometry of the Creutz ladder. The coordinates $j,\alpha$ label the sites, and $L$ is the length of the ladder. $J$ and $m$ are the corresponding hopping amplitudes. When a magnetic flux of $\Phi$ threads each plaquette, the complex horizontal amplitudes cause a phase change in the wavefunction of $\pm\phi/2=\pm \pi\Phi/\Phi_0$. b) Topological phase diagram for the Creutz ladder. The phase $\phi$ is proportional to the magnetic flux through a plaquette and $m$ is the vertical hopping amplitude. The left and right edges of the diagram represent the same systems. Two trivial ($\nu=0$) and two nontrivial ($\nu=\pm 1$) topological phases can be distinguished. c) Image of the function $(d_x(k),d_z(k))$ [Equation (\ref{eq:dCreutz})] along the Brillouin zone for some of the points in the phase diagram. Its winding number around the origin corresponds to the topological invariant $\nu$ of the phase the point belongs to. For \textit{O} and \textit{T}, $\nu$ is not well defined, and the image of the function passes through the origin.} 
    \label{fig:LattPh}
\end{figure}

The Creutz ladder (Figure \ref{fig:LattPh}a) consists of two chains of sites connected with horizontal, vertical and diagonal hopping amplitudes. Additionally, a magnetic field is applied perpendicularly to the plane of the ladder. This causes some hopping amplitudes to acquire a so-called Peierls phase and become complex. We choose the gauge in which the horizontal hoppings cause a change of phase of $e^{\pm i\phi/2}$ in the wavefunction, and the rest are real. The sign of the phase change is positive if the particle moves to the right (left) on the upper (lower) leg of the ladder, and negative if it moves in the opposite direction. For $m=0$ (no ``rung'' hopping terms), the lattice is bipartite, given that all first neighbours of sites with an odd $j$-coordinate correspond to an even $j$, and vice versa.

We will first consider the single-particle case, and go on to consider interactions
in Sec. \ref{sec:Hubbard}.
Let $c^{(\dagger)}_{j,\alpha}$ be the destruction (creation) operator for a boson in site $\ket{j,\alpha}$ of the ladder, where $j=1,...,L$ labels the rungs along the ladder and $\alpha=A,B$ labels the two legs. The Creutz Hamiltonian then takes the form:
\begin{align} \label{eq:CreutzH}
    \mathcal{H}_C=&-\sum_{j, \alpha}\left[ J_\alpha c_{j+1,\alpha}^\dagger c_{j,\alpha}+J c_{j+1,\alpha}^\dagger c_{j,\overline{\alpha}}\,+\right.\nonumber\\
    &+\left.\frac{m}{2}c_{j,\alpha}^\dagger c_{j,\overline{\alpha}}+H.c.\right]
\end{align}
where the horizontal hopping terms are $J_\alpha\!=\!J e^{i \sigma_\alpha \phi/2}$, with $\sigma_{A/B}\!=\!\pm 1$ and $\overline{A}=B$ and vice versa.
The model is thus specified by two hopping amplitudes
($J$ and $m$), and the magnetic flux, represented by $\phi$.

\begin{figure}[!htbp]
    \centering
    \includegraphics[width=\linewidth,trim=8 8 8 30,clip]{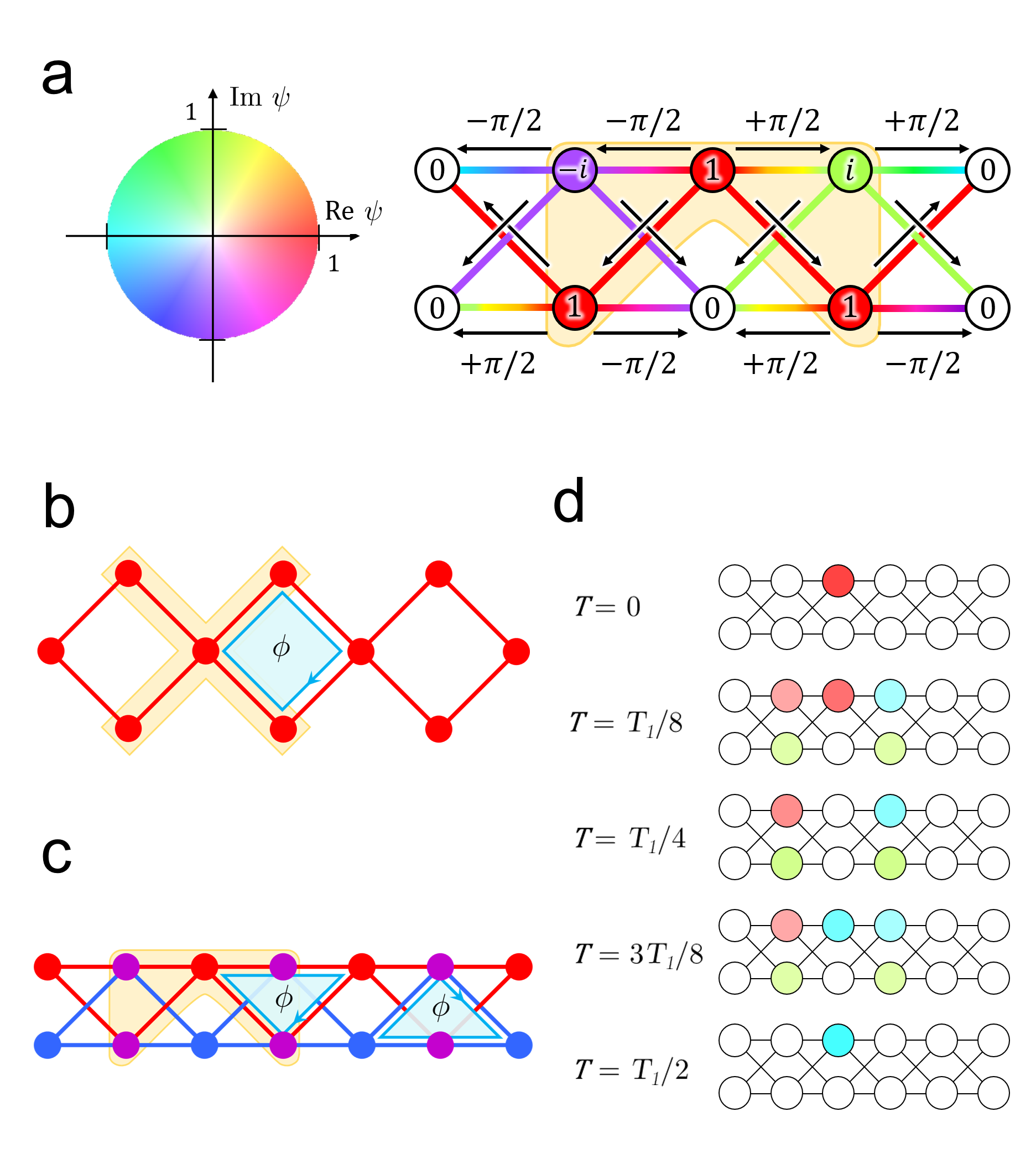}
    \caption{a) AB caging. In the system with $\phi=\pi$, the time evolution of the state $\ket{j,\alpha}$ never allows the particle to leave the AB cage centered on that site (shaded in orange). The numbers inside each site and the color code used show the phase acquired by the wavefunction when tunneling to the second neighbors of $j,\alpha$ along the two simplest (i.e. two-step) paths. Each pair of paths interfere destructively, and the particle never leaves the cage. The color code used to represent the phases of the wavefunction is specified by the color wheel on the left. This code is used throughout this work to represent complex quantities. b) Rhombus chain. A magnetic flux represented by the phase $\phi$ threads each rhombus. The AB cage of a spinal site is shaded in orange. c) A Creutz ladder with $m=0$ can be obtained by identifying the corresponding top and bottom sites of two rhombus chains, represented in red and blue. The purple sites mark the sites where the identification has taken place. The AB cages for this lattice are inherited from the rhombus chains that form it. d) Numerical simulation of the time evolution for state $\ket{3,A}$ (see Figure \ref{fig:LattPh}a). Half a period is shown. The wavefunction exhibits the oscillatory behaviour predicted in (\ref{eq:psit}). The color code in a) is also used here.}
    \label{fig:Cage}
\end{figure}

The physical magnetic flux that threads each plaquette, $\Phi$, is proportional to the total phase $\phi$ acquired by a particle moving along its border in a clockwise sense: $\phi=2\pi \frac{\Phi}{\Phi_0}=\frac{q \Phi}{\hbar}$, where $\Phi_0$ is the magnetic flux quantum, and $q$ is the electric charge of a particle. In the $m=0$ case, there are no rung-like bonds and plaquette-like paths cannot be defined. However, any 4-step paralelogram-shaped or triangular-shaped path is threaded by the same flux as a plaquette, and can be used instead (compare Figure \ref{fig:LattPh}a and Figure \ref{fig:Cage}c).

Since the unit cell of the system is composed of two sites, its band Hamiltonian is a two-dimensional Hermitian
matrix. Consequently, in momentum space it can be written in terms of the Pauli matrices as $\sH(k)=d_0 \mathds{1} + \vec d \cdot \vec \sigma$, where:
\begin{align}
    &d_0=-2J\cos k \cos \frac{\phi}{2} \\
    &\vec d = (-m-2J\cos k,\, 0,\, 2J\sin k \sin \frac{\phi}{2}). \label{eq:dCreutz}
\end{align}
In the abstract space formed by $\tilde{d}=(d_z,d_x,d_0)$, the $d_0$ axis corresponds to the possible band degeneracies (i.e. gap closings) in the bands of the Creutz Hamiltonian. This means that all Hamiltonians that satisfy $d_x(k_0)=d_z(k_0)=0$ for some $k_0$ in the first Brillouin zone (BZ) are metallic, with their two bands touching at all possible values of $k_0$. For the rest of the systems, the winding number of their directed path $\tilde{d}(k)$ (which is closed because of the periodicity of the BZ) around the $d_0$ axis determines the topological invariant of the Hamiltonian, of type $\mathds{Z}$, which serves to distinguish the different topological phases of the system. In Figure \ref{fig:LattPh}c, all paths have been projected down to the $(d_z(k),d_x(k))$ plane for simplicity, and thus the topological invariant corresponds to its winding number around the origin.

The relevant symmetries of the system, which protect the topological phases, \cite{Li2015,Jafari2019} are the spatial inversion symmetry $\sI:\, \sH(k)\to \sigma_x \sH(-k)$ (for all values of $\phi$), and the chiral symmetry $\sX=\sigma_y$\ for $\phi=\pm \pi$. Some authors consider the $\pi$-flux Creutz ladder to be in the symmetry class BDI, \cite{Altland1997} arguing that, in addition to the chiral symmetry, there are particle-hole and time reversal symmetries that are independently conserved \cite{Tovmasyan2013,Piga2017,Gholizadeh2018,Velasco2019}. Other authors classify it as AIII, claiming that the magnetic flux breaks time reversal symmetry. \cite{Li2015,Jafari2019}  For other values of $\phi$, however, only inversion symmetry protects the topological phases, and is not one of the traditional symmetries considered in the classification of topological insulators. \cite{Altland1997} The topological systems in which it plays a significant role cannot be sorted into the usual symmetry classes, and do not usually exhibit edge states (something unusual for a topological insulator). \cite{Hughes2011} The Creutz ladder is an unusual example, as it does present edge states protected by $\sI$, making it an excellent tool to probe these exotic symmetry classes. It is also an outlier in the sense that all 1D topological insulators with a $\mathds{Z}$-type invariant in the traditional classes require chiral symmetry, while it is absent in the Creutz ladder for most values of $\phi$.

We show the topological phase diagram of the system as a function of the phase induced by the magnetic flux, $\phi$, and the vertical hopping amplitude, $m$, in Figure \ref{fig:LattPh}b. \cite{Li2015} Four topological phases can be distinguished. The system is topologically trivial for $|m|>2J$, with a vanishing winding number, and thus a Zak phase of zero. For $|m|<2J$, two different topological phases exist, with winding numbers of $\nu=\pm 1$, depending on the value of $\phi$. Both phases are associated with a Zak phase of $\sZ=\pi$, given that this quantity is defined modulo $2\pi$ (see Appendix \ref{app:Zak}). The boundaries of these regions correspond to metallic systems, in which the two bands of the model touch. The phase diagram is periodic in $\phi$, with period $4\pi$. This means that the left and right sides of Figure \ref{fig:LattPh}b represent the same points in phase space. As a simple consistency check, we calculated the Zak phase numerically for different points of the phase space using Equation \ref{eq:ZakComp}.

\subsection{Aharonov-Bohm localization and topological edge states}
The Creutz model with $\phi=\pm \pi, \, m=0$ 
is of particular interest because it presents non-dispersive energy bands. Such flat bands have a constant dispersion relation $E_\pm(k) = \pm 2J$, and thus a vanishing group velocity, meaning that particles stay localized in space. This effect is caused by interference arising from Aharonov-Bohm phases, and has been termed AB caging. 
This effect is most clearly seen in the rhombus chain lattice, shown
in Figure \ref{fig:Cage}b). Consider a particle initially localized on one of
the central sites with a coordination number of four, which we can term
a spinal site. It will propagate to a neighboring spinal site in two successive
tunneling processes, and two different routes are available for this; either
via the upper side of the chain, or the lower side. Due to the Peierls phases
acquired along each path,
there is a phase difference of $\phi$ between these paths, and consequently
if $\phi = \pi$, the two paths interfere destructively. In that case the occupation
of the neighboring spinal sites is frozen at zero, and the particle only undergoes
a breathing motion from the initial site to its four neighbours. These five sites
constitute the particle's ``cage''.

Being a quantum interference effect, this phenomenon requires no disorder, and indeed, has been shown to be robust to moderate levels of it. \cite{Kuno2019} As well as the rhombus chain lattice,
\cite{Vidal2000,Creffield2010,Movilla2011} this caging occurs in a variety of other flat-band bipartite lattices, including the $\sT_3$ or dice lattice \cite{Vidal1998a} and the Lieb lattice. \cite{lieb} The most direct way to observe AB caging is to measure the ocupation numbers for all sites in the lattice during the time evolution, and check that a particle initially localized in a single site only visits a small number of them (its cage). This approach is convenient for waveguide lattice setups, given that the initial state requires only pumping light into a single site. Experimental observation of AB caging has indeed been recently observed in this way for a waveguide rhombus chain using classical light. \cite{Mukherjee2018}

The Creutz ladder differs from the rhombus chain, in that every site has the same
coordination number. Their geometries, however, are closely related.
A Creutz ladder with no rungs ($m=0$) can be obtained by identifying the top and bottom sites of two rhombus chains (see Figure \ref{fig:Cage}c). Accordingly we would expect
the resultant Creutz ladder to inherit the AB cages from the constituent
rhombus chains. 
This expectation can be verified explicitly by analyzing the Creutz ladder
using a basis of Wannier states $\ket{j\pm}$ with support on the four sites ($j,A;j,B;j+1,A;j+1,B$) that form each one of the plaquettes in the ladder, \cite{Tovmasyan2013,Piga2017,Kuno2019} defined by the following expression:
\begin{equation}
    \ket{j\pm} = \frac{1}{2}\big[\ket{j,A}+i\ket{j,B}\mp i\ket{j+1,A}\mp\ket{j+1,B}\big]. \label{eq:wan}
\end{equation}
Their energies in the flat band limit are $E_{\pm}=\pm 2J$. Any localized state with no support on the ends of the system can be expressed as a superposition of these plaquette states. In particular,
\begin{equation}
    \ket{j,\alpha} = \frac{1}{2}\big[-i\ket{j-}+i\ket{j+}+\ket{j\!+\!1\, +}+\ket{j\!+\!1\, -} \big].
\end{equation}
Thus, in this limit the time evolution of state $\ket{j,\alpha}$ will be periodic, with a frequency equal to the absolute value of the energy of the bands. A similar analysis can be done for the eigenstates of the rhombus chain model \cite{Vidal2000}, although it is less intuitive than the discussion in terms of AB phases we give above. In the Creutz laddder, using the definition of the Wannier basis [Equation (\ref{eq:wan})] in the time-dependent state yields:
\begin{align}
    &\ket{\psi (t)} = \frac{1}{2}\sin (2J t)\big[\! -\! \sigma_\alpha \ket{j + 1,\alpha} \! + \! i\ket{j + 1,\overline\alpha}  \nonumber\\
    &+\! \sigma_\alpha \ket{j - 1,\alpha} \! +\!  i\ket{j - 1,\overline\alpha}   \big] + \cos (2J t)\ket{j,\alpha}, \label{eq:psit}
\end{align}
which describes an oscillatory behaviour between the original site and its first neighbours. Here, $J$ is the diagonal hopping amplitude and $t$ is time. As expected, this behaviour is reminiscent of the one found in the rhombus chain. \cite{Vidal2000} The initial site and its first neighbours constitute the AB cage of the particle. Site $j,\overline \alpha$, in the same rung of the ladder as the initial site, is one of its second neighbours if $m=0$ and, as such, is not included in the cage. An exact numerical simulation was carried out for a system with $L=6$. The time step used was $\Delta t=0.001$, where time is measured in units of $J^{-1}$ (taking $\hbar=1$).
The results are shown in Figure \ref{fig:Cage}d. The obtained half-period was $T_1/2=1.572$. This result agrees with the analytical calculation in Equation (\ref{eq:psit}), where $T_1=2\pi/2 J=\pi$. A similar analysis shows that the cage for a particle localized at a single end site (e.g. $1,A$, see Figure \ref{fig:LattPh}a) is constituted by the four sites closer to that end of the ladder (for $1,A$, these are $1,A;\, 1,B;\, 2,A;\, 2,B$). 

\begin{figure}[!htbp]
    \centering
    \includegraphics[width=\linewidth,trim=15 8 15 30 ,clip]{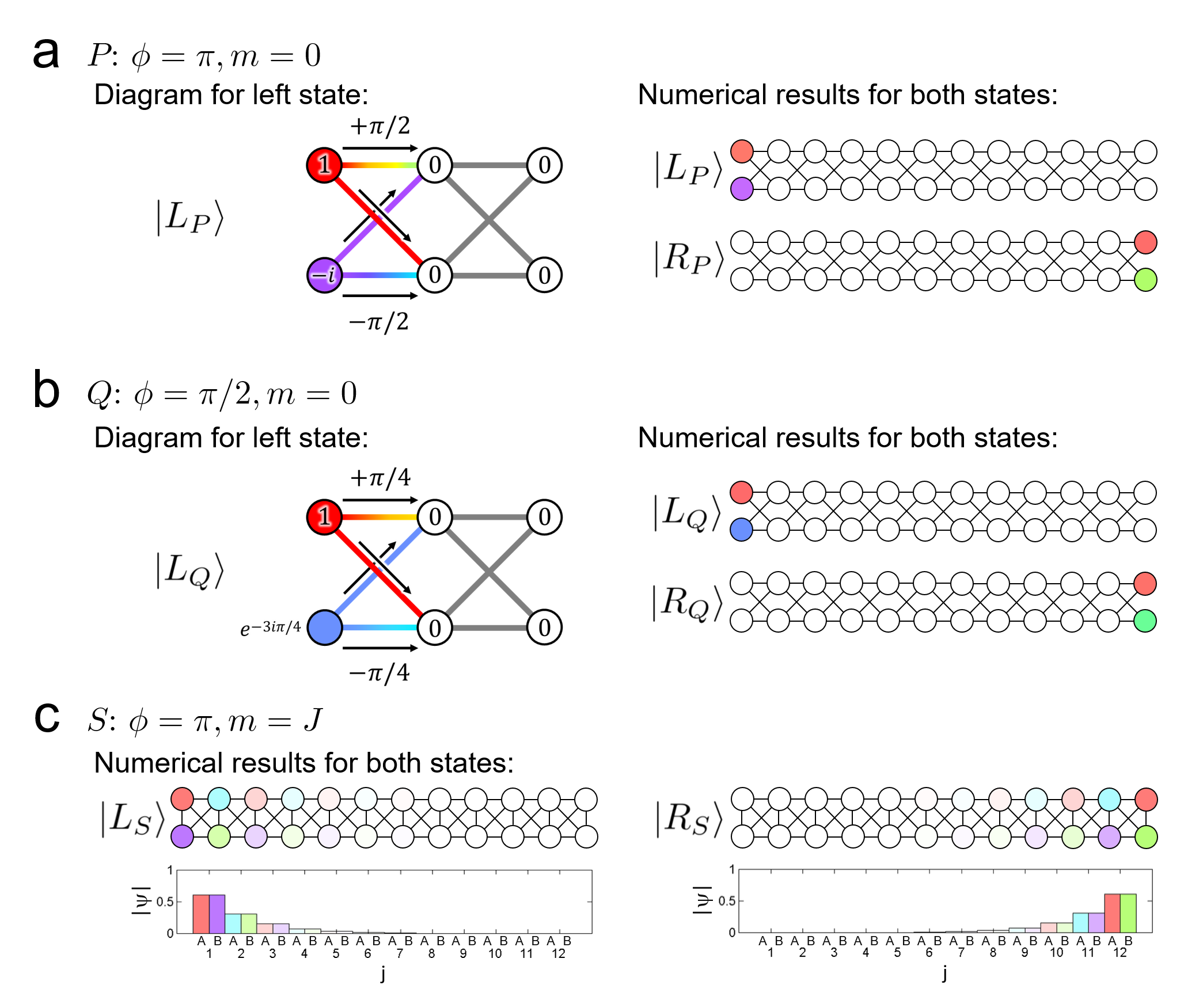}
    \caption{Wavefunctions for both edge states corresponding to the systems labeled a) \textit{P}, b) \textit{Q} and c) \textit{S} in the $\nu=1$ phase (see Figure \ref{fig:LattPh}b) for a ladder of length 12, obtained using numerical exact diagonalization on the Creutz Hamiltonian. In systems \textit{P} and \textit{Q}, AB interference keeps the particle localized in the two end sites. Diagrams similar to Figure \ref{fig:Cage}a are included (left). In system \textit{S}, localization is  exponential. Bar graphs have been included in case \textit{S} to illustrate the exponential decay of the wavefunctions. In all diagrams, the color of the sites and bars correspond to the phase and modulus of the wavefunction, following the color code specified in Figure \ref{fig:Cage}a.}
    \label{fig:3}
\end{figure}

In the case with $m\ne 0$, the connectivity of the ladder changes, preventing it from being locally bipartite (the rungless ladder is not globally bipartite with periodic boundary conditions if $L$ is odd) and allowing new paths that do not interfere destructively and release the particle from its cage. This prevents AB caging from taking place.

Another interesting feature of the Creutz model is its edge states, only present in the topological phases and protected by the aforementioned symmetries. A simple yet partially unexplored question so far is the appearance of these edge states for different values of the parameters of the system. The form of the edge states depends drastically on the presence or absence of the vertical hopping amplitude. For $m=0$, AB interference localizes the particle entirely in the first or last sites of the ladder, as was already pointed out by Creutz for the $\pi$-flux case. \cite{Creutz1999} As an example, in the states $\ket{L_P}$ and $\ket{L_Q}$, the paths $\ket{1A}\to\ket{2A}$ and $\ket{1B}\to\ket{2A}$ cancel each other out (see Figures \ref{fig:3}a,b). Any other, more complicated, path to site $\ket{2A}$ from one of the end sites $\ket{1\alpha}$ gets canceled out with the corresponding one from site $\ket{1\overline\alpha}$. All edge states for $m=0$ obey the formula:
\begin{align}
    &\ket{L_\phi}=\frac{1}{\sqrt{2}}\left[\ket{1,A}-e^{i\phi/2}\ket{1,B}\right] \label{eq:Ledge}\\
    &\ket{R_\phi}=\frac{1}{\sqrt{2}}\left[\ket{L,A}-e^{-i\phi/2}\ket{L,B}\right]. \label{eq:Redge}
\end{align}
In particular, the edge states of the systems corresponding to the points \textit{P} and \textit{Q} in the topological phase diagram (Figure \ref{fig:LattPh}b) have the expression:
\begin{align}
    &\ket{L_P}\equiv\ket{L_{\phi=\pi}}=\frac{1}{\sqrt{2}}\left[\ket{1,A}-i\ket{1,B}\right] \nonumber\\
    &\ket{R_P}\equiv\ket{R_{\phi=\pi}}=\frac{1}{\sqrt{2}}\left[\ket{L,A}+i\ket{L,B}\right] \nonumber\\
    &\ket{L_Q}\equiv\ket{L_{\phi=\pi/2}}=\frac{1}{\sqrt{2}}\left[\ket{1,A}+e^{3i\pi/4}\ket{1,B}\right] \nonumber\\
    &\ket{R_Q}\equiv\ket{R_{\phi=\pi/2}}=\frac{1}{\sqrt{2}}\left[\ket{L,A}+e^{-3i\pi/4}\ket{L,B}\right]. \nonumber
\end{align}
This is confirmed by the numerical results shown in Figure \ref{fig:3}a,b.

For all values of $\phi$, the particle is AB-localized in only two sites as long as $m=0$, a remarkable result given that the AB caging for single-particle bulk states only occurs for the flat-band systems $m=0$, $\phi=\pm\pi$. At this point, it is important to have in mind that the details of the phase differences of the wavefunction between the different sites are gauge dependent, but the physical conclusions obtained (such as localization) are not. Adding rungs to the ladder creates many new paths with amplitudes that do not vanish when added together. Similarly to the case of AB caging, 
these changes cause the wavefunction to leak out of the end sites, and present an exponential profile instead, as is frequently seen in physical boundary states. A bar graph illustrates the exponential nature of the states in Figure \ref{fig:3}c. The edge states represented in Figure \ref{fig:3}a-c correspond to the systems labeled \textit{P}, \textit{Q} and \textit{S} in the phase diagram from Figure \ref{fig:LattPh}b, and were obtained diagonalizing the Creutz Hamiltonian numerically for a length of $L=12$.

\subsection{Photonic implementation}

A direct implementation of a physical Creutz-like structure will encounter the problem of designing the diagonal crosslinks. Some efforts have been made in this direction. For example, an ultracold atom setup was proposed in which two overlapping zigzag optical lattices could be used to realize the lattice geometry. \cite{Sun2017}

An alternative way to overcome this challenge, which we will explore in this work, is to use a synthetic dimension. Synthetic dimensions are widely used in photonic materials \cite{Yuan2018} and ultracold atom systems \cite{Boada2012,Celi2014} to increase the dimensionality of the lattice. A synthetic dimension employs a non-spatial discrete degree of freedom of the system (often the frequency mode number in photonic arrays and the different atomic hyperfine levels in ultracold atoms) to increase the number of dimensions by one. In this way the physics of a system with $d+1$ spatial dimensions can be simulated by a $d$-dimensional system with one additional synthetic dimension.

\begin{figure}[!htbp]
    \centering
   \includegraphics[width=\linewidth,trim=108 8 108 8,clip]{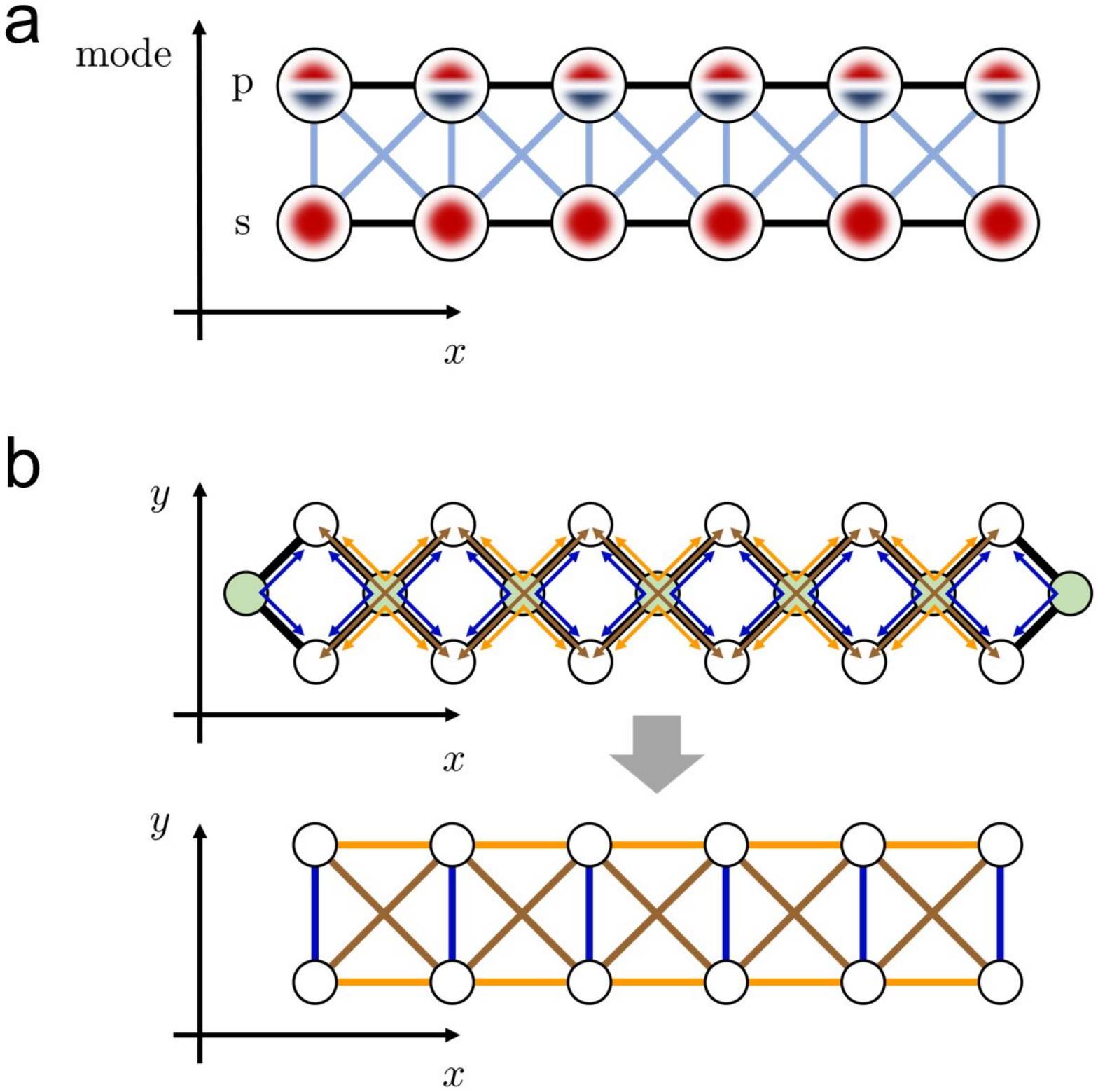}
    \caption{a) Creutz ladder in one spatial and one synthetic dimension. In a photonic setup two different electromagnetic modes would represent the two legs. In a cold atom system, \textit{s} and \textit{p} atomic orbitals would be used instead. This setup was implemented recently using ytterbium atoms in an optical lattice. \cite{Kang2019}
    b) Simulation of the Creutz ladder using a rhombus chain with a large energy offset in the sites with four first neighbours (shaded in green). The particles are restricted to the top and bottom sites, but they can tunnel from one site to another via virtual processes involving the middle sites. This way, an effective lattice with the geometry of the Creutz ladder is obtained. The synthetic magnetic field is inherited from the one in the rhombus chain. \cite{Mukherjee2018}}
    \label{fig:4}
\end{figure}

The simplest way to implement the Creutz ladder in this manner in a photonic system would be to use a two-level degree of freedom to simulate the vertical dimension of the ladder (i.e. the two different legs) in a one-dimensional chain of resonators. This way, the diagonal links exist in the synthetic space, and their crossing poses no practical problem. One possible realization using a 1D array of parallel waveguides is to make use of two different electromagnetic modes to represent the two legs of the ladder in each waveguide, as shown in Figure \ref{fig:4}a). This same scheme, but using atomic orbitals instead of electromagnetic modes, was recently used in an ultracold ytterbium atom system in an optical lattice to achieve one of the first experimental implementations of the Creutz ladder. \cite{Kang2019} A system very similar to the Creutz ladder was also previously realized in cold atoms using the same approach. \cite{Li2013}

Another way to tackle the problem would be to simulate the Creutz ladder in a different system
using the approach suggested in Ref. \cite{Mukherjee2018}, which would
make use of the topological similarities between the Creutz ladder and the rhombus chain. In that work, a photonic rhombus chain was constructed using a waveguide lattice. No synthetic dimensions are necessary to build the rhombus lattice, as no tunneling processes cross, making its physical realization considerably more straightforward. If a sufficiently high potential energy is then imposed on the spinal sites, the edge sites will then behave effectively as the two legs of a Creutz ladder (Figure \ref{fig:4}b). 
A very similar photonic rhombus chain was also realized in another recent work. \cite{Kremer2018}
The synthetic magnetic field can be obtained with auxiliary waveguides that retard the wavefunction appropriately, \cite{Kremer2018} or with Floquet engineering, combining a periodic modulation of the refraction index with an energy gradient. \cite{Mukherjee2018} In these systems, it is especially easy to observe the AB caging. In the scalar-paraxial approximation used, the direction along the waveguides behaves as the time dimension of the simulated model. That way, if light is injected into the waveguide $j,\alpha$, the oscillatory movement predicted by Equation (\ref{eq:psit}) will be observed as spatially periodic (\textit{breathing}) movements in and out of its first neighbours. It would thus be possible to obtain the Creutz ladder analogue of these results (i.e. Figure \ref{fig:Cage}d) in the existing photonic implementations of rhombus chain, using the mapping between them detailed above.

The highly localized nature of the edge states of the Creutz ladder implies that they too can be directly observed in a similar way.
Distinguishing them, however, is inherently harder than simply observing their localization, because of the phase difference needed between different sites even in the simplest edge states [Equations (\ref{eq:Ledge},\ref{eq:Redge})].

The discussion above shows that the noninteracting Creutz ladder can be implemented experimentally in photonic systems, and that interesting new phenomena are already accessible in that framework. In the following section, we will discuss the considerably more complicated question of including on-site interactions in the Creutz model, and its experimental feasibility.

%SEC CREUTZ-HUBBARD
\section{Creutz-Hubbard ladder}
\label{sec:Hubbard}

As we noted earlier, interactions between particles play an interesting role in both topological and flat band systems. This motivates the inclusion of an interaction term in the Creutz Hamiltonian. In the literature, nearest neighbor \cite{Piga2017,Sticlet2014, Kuno2019} or on-site interactions, both attractive \cite{Tovmasyan2016,Tovmasyan2018} and repulsive \cite{Takayoshi2013,Tovmasyan2013} have been considered, We will consider an on-site repulsive ($U>0$) Hubbard interaction term in the Creutz Hamiltonian:
\begin{equation} \label{eq:CHubbH}
    \sH_{C\!H}=\sH_C + \frac{U}{2}\sum_{j,\alpha} c_{j\alpha}^\dagger c_{j\alpha}^\dagger c_{j\alpha} c_{j\alpha}.
\end{equation}

Given that $c^{(\dagger)}_{j\alpha}$ are bosonic operators, there is no limit to the number of particles that a single site can hold. Nevertheless, for reasons that will become apparent soon, we will focus on the case where only two particles populate the ladder, the minimal case for interaction effects to occur. We are able to do this because the Creutz-Hubbard Hamiltonian preserves the number of particles. Additionally, its correlated nature prevents the use of a single-particle approach. We will use a bosonic Fock space basis $\{\ket{1_{i,\alpha}1_{j,\beta}}\}_{i,\alpha,j,\beta}$, defined for the two-particle subspace as:
\begin{equation}
    \begin{cases}
    &\!\!\!\!\!\!\ket{1_{i,\alpha}1_{j,\beta}} \equiv c^\dagger_{i,\alpha} c^\dagger_{j,\beta}\ket{0} \kern 15 pt  i\ne j  \mathrm{ \ or \ } \alpha \ne \beta \\
    &\!\!\!\!\!\!\ket{1_{i,\alpha}1_{i,\alpha}} \equiv \ket{2_{i,\alpha}} \equiv \frac{1}{\sqrt{2}} (c^\dagger_{i,\alpha})^2 \ket{0} .
    \end{cases}
\end{equation}
The nature of the photonic implementation we will consider in this work forces us to also consider a first quantization Hilbert basis of the two-particle space $\{\ket{i,\alpha;j,\beta}\}_{i,\alpha,j,\beta}$. The main difference between both bases is that the states in the former are intrinsically symmetric under the exchange of both particles (and so $\ket{1_{i,\alpha}1_{j,\beta}}$ is the same state as $\ket{1_{j,\beta}1_{i,\alpha}}$), while in the latter the states $\ket{i,\alpha;j,\beta}$ and $\ket{j,\beta;i,\alpha}$ are distinct. The relationship between both bases is: $\ket{1_{i,\alpha}1_{j,\beta}}=[\ket{i,\alpha;j,\beta}+\ket{i,\alpha;j,\beta}]/\sqrt{2};\, \ket{2_{i,\alpha}}=\ket{i,\alpha;i,\alpha}$

Using each of these basis, any state $\ket{\psi}$ can be expressed as:
\begin{equation} \label{eq:bases}
    \ket{\psi} =  \sum_{(i,\alpha)\leq(j,\beta)} \eta_{i,\alpha;j,\beta} \ket{1_{i,\alpha}1_{j,\beta}} = \sum_{\forall i,\alpha,j,\beta} \lambda_{i,\alpha;j,\beta}\ket{i,\alpha;j,\beta},
\end{equation}
where we have defined an ordering for the sites $i,\alpha$: $1A<1B<\cdots<LA<LB$. The bosonic nature of the particles imposes the constraint $\lambda_{i,\alpha;j,\beta}=\lambda_{j,\beta;i,\alpha}$, and the components in each basis are related by $\eta_{i,\alpha;j,\beta}=[\sqrt{2}+\delta_{i,j}\delta_{\alpha,\beta}(1-\sqrt{2})]\lambda_{i,\alpha;j,\beta}$. We will expand on the need for the two different bases in section \ref{subsec:photo}.

\subsection{Effective Hamiltonian, doublon caging and edge states}
For sufficiently large values of $U$, the repulsive interaction can actually bind both particles together. This can be understood in energetics terms: if $U \gg J,m$, the states of the system with a doubly-occupied site have a much higher energy than the others. As there is no dissipation in our model, conservation of energy means that a doubly occupied state cannot evolve into a pair of separated particles under time evolution. Both particles thus move together around the lattice as a pair, forming a \textit{doublon}. 
Thus, a doublon is a quasiparticle composed of two particles which are bound as a result of a repulsive Hubbard interaction, in which both particles occupy the same site. A model where dissipation is taken into account would see doublons break apart after a certain lifetime, due to different mechanisms as doublon-doublon scattering or phonon scattering. \cite{Bello2017} 

The fact that the doublon bands (formed by all the doubly occupied states of the Fock basis) are decoupled from the rest of the spectrum (the singly occupied states) can be exploited to execute a Schrieffer-Wolff transformation and obtain an effective Hamiltonian for the doublon subspace. \cite{Bravyi2011,Hofmann2012,Bello2016} The resulting Hamiltonian, for a finite system, takes the form:
\begin{align} \label{eq:Heff}
    \mathcal{H}_{\mathrm{eff}}&=\sum_{j, \alpha}\left[ \frac{J^2_\alpha}{U} d_{j+1,\alpha}^\dagger d_{j,\alpha}+\frac{J^2}{U} d_{j+1,\alpha}^\dagger d_{j,\overline{\alpha}}\,+\right.\nonumber\\
    &+\left.\frac{m^2}{2U}d_{j,\alpha}^\dagger d_{j,\overline{\alpha}}+H.c.\right]+\frac{U}{2}\sum_{j,\alpha}d_{j\alpha}^\dagger d_{j\alpha} +\nonumber\\
    &-\sum_{\alpha}\mu[d_{1\alpha}^\dagger d_{1\alpha} + d_{L\alpha}^\dagger d_{L\alpha}]   +\Delta
\end{align}
where $d^{(\dagger)}_{j\alpha}\equiv (c^{(\dagger)}_{j\alpha})^2$ are the doublon creation and annihilation operators, and $\Delta=[(2m^2+8J^2)/U]\mathds{1}$ is an energy offset of the whole Hamiltonian with respect to the original scale of energy. The first sum is a renormalized Creutz model, with $-J\to J^2/U,\, -J_\alpha\to J^2_\alpha/U,\, -m \to m^2/U$. The second term is the Hubbard interaction term, which remains unchanged. Additionally, a chemical potential $-\mu=-2J^2/U$ appears in the end sites, because of their different coordination number than the rest of the lattice. This same phenomenon was demonstrated in other 1D and 2D effective models for doublons. \cite{Bello2016,Bello20162D}

The rungless (i.e. $m=0$) $\pi$-flux Creutz-Hubbard ladder with repulsive interactions has been studied in the literature for different values of its filling factor. These studies have shown that, for a dense enough system, a Tomonaga-Luttinger liquid (TLL) of doublons is spontaneously formed, while some fraction of the particles remain unpaired and localized in the background. \cite{Takayoshi2013,Tovmasyan2013} The dynamics in this situation have some similarities and differences with the sparse system with one or two particles that we focus on. The most important common ground is that single particle localization for $\phi=\pi$ is due to AB caging for any filling factor. On the other hand, the interacting nature of the Hamiltonian makes the spectrum of the system fairly sensitive to the filling factor. Indeed, all states with double occupancy remain in the high energy region of the spectrum for sparse systems, but as the number of particles increases they eventually take a part in the ground state of the TLL. This illustrates the complexity and richness that often characterize doublon dynamics in Hubbard models.

Given that the doublon will have an electric charge $2q$ (with $q$ the charge of one particle), we can expect to find analogues of the behaviour found in the single particle dynamics for half the magnetic flux than in the latter case. In particular, AB caging for doublons is expected to be found for $\phi=\pm\pi/2$. Because of the periodicity of the phase diagram in $\phi$, caging should also be found for $\phi=\pm 3\pi/2$. For these values of the flux, the independent particles are not localized, and this paints an interesting picture: an ensemble of localized doublons and free-moving particles. This is the opposite situation to the rungless $\pi$-flux regime, in which particles were caged and pairs could propagate. The behaviour for $m\ne 0$ is analogous to the single-particle case: including a vertical hopping amplitude breaks AB caging for all values of $\phi$. In the following, we will focus on the case with $m=0$.

\begin{figure}[!htbp]
    \centering
    \includegraphics[width=\linewidth,trim=8 8 8 8,clip]{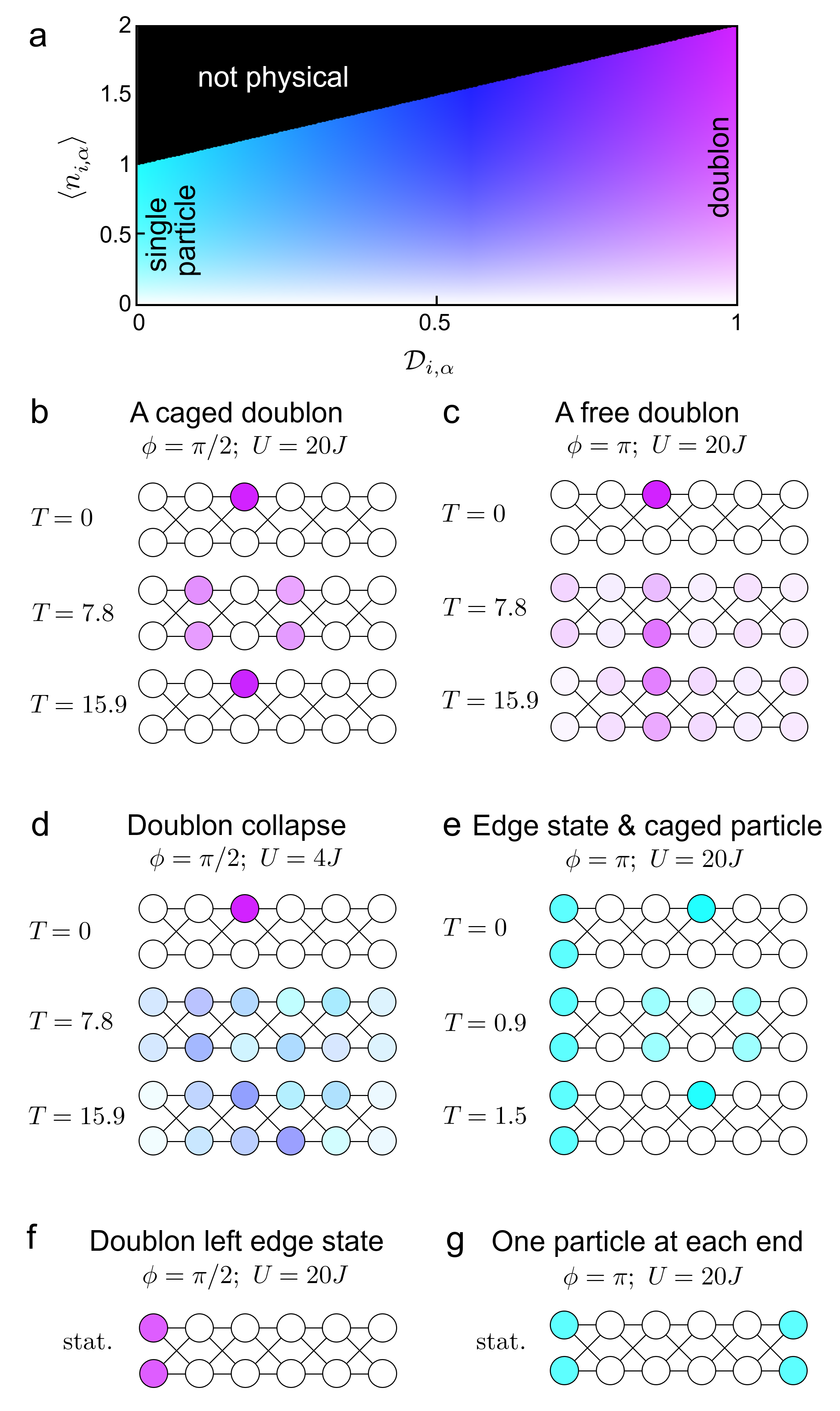}
    \caption{Numerical time evolution of the two-particle Creutz-Hubbard ladder. All simulations use $J=1$, $m=0$ and a time step of $\Delta t=0.3$. The left (right) column uses $\phi=\pi/2 \, (\pi)$ a) Color code used for the figures. $\langle n_{i,\alpha} \rangle$ is the expected value for the occupation number of each site [see Equation(\ref{eq:n})] and $\sD_{i,\alpha}$ is a measure of the \textit{doublonness} of the state [Equation (\ref{eq:d})]. b) A caged doublon around site $3,A$. This can be observed for $\phi=\pm\pi/2;\pm 3\pi/2$ and $U\gg J$. c) A doublon which is not localized, the behaviour for the rest of values of $\phi$ if $U\gg t$. d) A doublon collapsing into two quasi-independent particles. This happens if $U \simeq J$, for any value of $\phi$. e) A particle remains localized at the left edge and the other one is caged around site $4,A$. This can only happen for $\phi=\pm\pi$. Otherwise, the second particle would not be localized. f) Doublon left edge state $\ket{2L_\phi}$ [Equation (\ref{eq:2Ledge})], possible for any value of $\phi$ if $U\gg J$. This state is stationary. g) Each particle stays localized at one end. This can also happen for any value of $\phi$, and the state is also stationary.}
    \label{fig:5}
\end{figure}

Doublon edge states, which have been studied for different 1D and 2D systems \cite{Bello2016,Bello20162D,Gorlach2017,Salerno2018}, cannot be treated in the same way as their single particle counterparts in 1D and quasi-1D systems, as the usual bulk-boundary correspondence is not valid and the value of the Zak phase is not correlated to the existence of topological doublon edge states. \cite{Gorlach2017} Consequently a different approach must be employed. In the $m=0$ case, the simplicity of the AB-localized single-particle edge states in Equations (\ref{eq:Ledge},\ref{eq:Redge}) allows us to probe for analogous doublon edge states. The relative phase between the sites must now be twice that of the single-particle states, to match the AB phase acquired in each hopping by the doublon of charge $2q$:
\begin{align}
    &\ket{2L_\phi}=\frac{1}{\sqrt{2}}\left[\ket{2_{1,A}}-e^{i\phi}\ket{2_{1,B}}\right] \label{eq:2Ledge}\\
    &\ket{2R_\phi}=\frac{1}{\sqrt{2}}\left[\ket{2_{L,A}}-e^{-i\phi}\ket{2_{L,B}}\right]. \label{eq:2Redge}
\end{align}
Exact numerical time evolution of these states confirm their stationary nature and identifies them as eigenstates of the Hamiltonian (\ref{eq:CHubbH}) (see Figure \ref{fig:5}f). Photonic doublon edge states can be used to create maximally entangled multiphoton (so-called \textit{NOON}) states, which find applications in high-precision quantum metrology. \cite{Afek2010,Gorlach2017a} In our system, the state $\ket{2L_\phi}+\ket{2R_\phi}$ constitutes a NOON state. A driving potential could also be used to implement a doublon transfer protocol between both ends of the ladder, which could find applications in the field of quantum information. \cite{Bello2016}

Figure \ref{fig:5} shows several numerical simulations of the rungless Creutz-Hubbard ladder for different values of its parameters. To get an intuitive feeling for the dynamics, color was used to encode two important features of the wavefunctions (Figure \ref{fig:5}a): the expectation value of the occupation number for each site, $\langle n_{i,\alpha} \rangle$, and a measure for the \textit{doublonness} of the wavefunction, $\sD_{i,\alpha}$. This measure is the probability that, if the site $i,\alpha$ is measured and found not to be empty, two particles (and not only one) are found in it. It can be calculated as a conditional probability. The numerator represents the probability of finding a doublon in site $i,\alpha$ if a measure is performed, and the denominator is the probability of finding that the site is not empty:
\begin{align}
    &\langle n_{i,\alpha} \rangle = \bra{\psi}c_{i,\alpha}^\dagger c_{i,\alpha}\ket{\psi} =\nonumber   \\
    &=\sum_{j\ne i,\beta\ne \alpha} |\lambda_{i,\alpha;j,\beta}|^2 + 2|\lambda_{i,\alpha;i,\alpha}|^2 \label{eq:n}\\
    &\sD_{i,\alpha} = \frac{|\lambda_{i,\alpha;i,\alpha}|^2}{\sum_{\forall j,\beta} |\lambda_{i,\alpha;j,\beta}|^2}. \label{eq:d}
\end{align}

\begin{figure*}[!htbp]
    \centering
    \includegraphics[width=\linewidth,trim=17 112 8 98,clip]{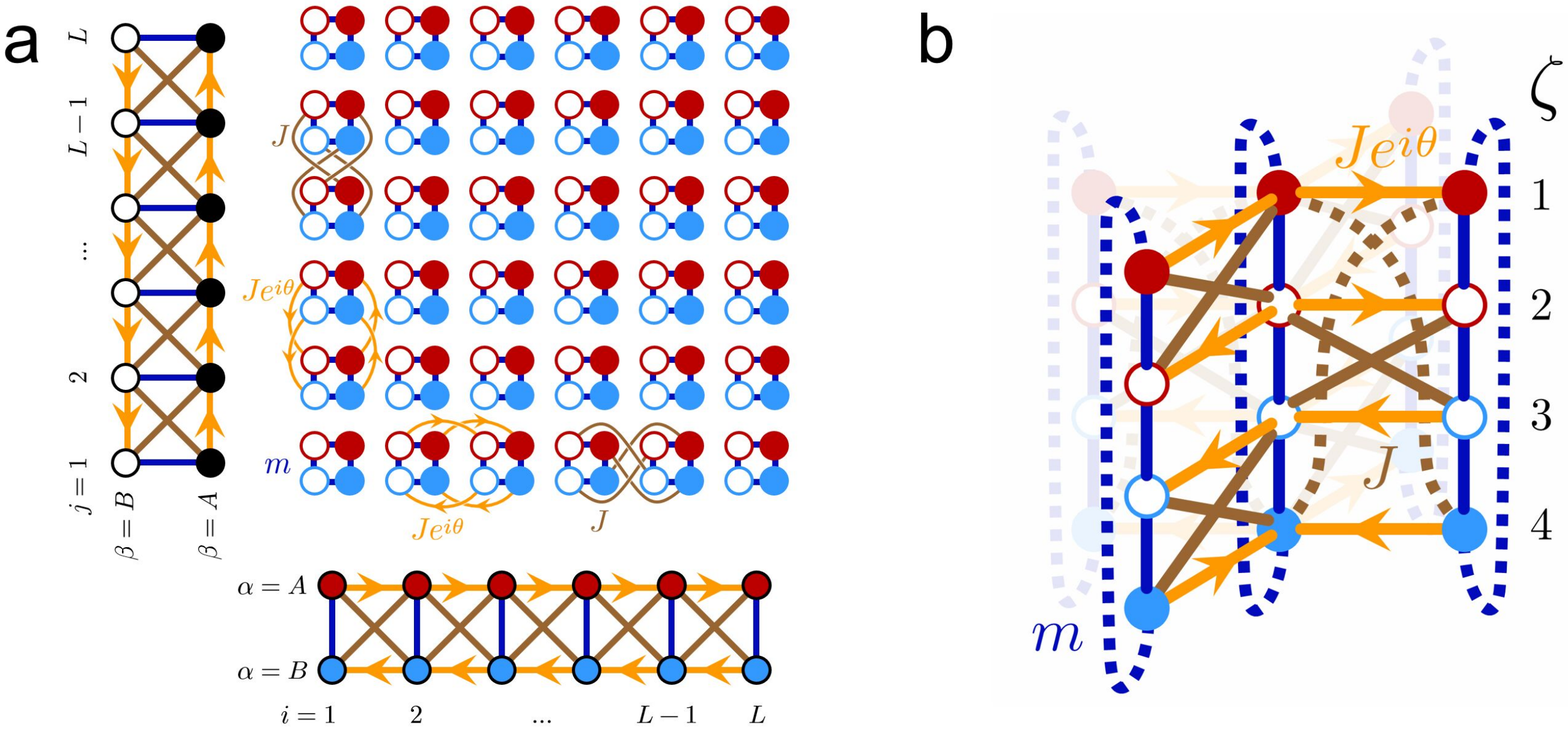}
    \caption{a) Quasi-2D lattice for the effective model, result of the Cartesian product of two Creutz ladders (left and bottom). All of the intercell (orange and brown) hoppings in the figure must be established between all adjacent unit cells. b) Portion of the analogous lattice in 3D. The three nonlocal hoppings per unit cell are represented by dotted lines.}
    \label{fig:6}
\end{figure*}

The initial state for Figures \ref{fig:5}b-d is $\ket{2_{3,A}}$, in which both particles start at site $3,A$. The initial state for Figure \ref{fig:5}e is $\ket{L_P}\otimes\ket{4,A}=(1/\sqrt{2})[\ket{1_{1,A}1_{4,A}}-i\ket{1_{1,B}1_{4,A}}]$ [see Equation (\ref{eq:Ledge})], for \ref{fig:5}f it is $\ket{2L_{\phi=\pi/2}}$ [Equation (\ref{eq:2Ledge})] and for \ref{fig:5}g, $\ket{L_P}\otimes\ket{R_P}=(1/2)[\ket{1_{1,A}1_{L,A}}-i\ket{1_{1,B}1_{L,A}}+i\ket{1_{1,A}1_{L,B}}+\ket{1_{1,B}1_{L,B}}]$. As shown in the figure, the characteristic times for doublon dynamics are around one order of magnitude greater than the ones for single particle dynamics.

\subsection{Bidimensional model and photonic implementation} \label{subsec:photo}

The classical light propagating in a photonic lattice, in the paraxial approximation, follows a wave equation that can be mapped to the Schr\"{o}dinger equation, and thus the amplitude of the electric field can be regarded as analogous to a wavefunction in (first quantization) quantum mechanics. This is the analogy on which quantum simulation in photonic lattices is based. To implement the first quantization Hilbert space corresponding to the two-particle 1D system, a 2D system is needed. This is a well-known method of treating the degrees of freedom in low-dimensional systems, and has been employed for both conventional \cite{Longhi2011,Longhi2011a,Mukherjee2016} and topological materials. \cite{Liberto2016,Gorlach2017}.

Using the Fock basis, the Schr\"odinger equation for the two-particle photonic Creutz-Hubbard system, $\sH_{C\! H}\ket{\psi}=E\ket{\psi}$, turns into a system of linear equations that can be interpreted as describing a single bosonic particle hopping around a quasi-2D lattice. This system of equations is calculated in Appendix \ref{app:Syst}. The resulting lattice can be obtained as the Cartesian product of two Creutz ladders.
The sites in this new lattice are labeled with the four indices $i,\alpha,j,\beta$, where $i,\alpha$ would be the position of the first particle in the original model, and $j, \beta$ that of the second one (see Figure \ref{fig:6}a).
In this way, the diagonal sites $i,\alpha,i,\alpha$ correspond to states with a doubly occupied site in the original Creutz lattice. If the particle never leaves the diagonal in the 2D lattice, that implies that the corresponding 1D doublon would not decay into two separate particles. The interaction energy $U$ has to be reinterpreted as an on-site potential in the diagonal sites of the 2D model.
For the photonic lattice to correctly emulate a bosonic system, all initial states considered must be symmetric under exchange of the two particles (this is equivalent to the aforementioned constraint $\lambda_{i,\alpha;j,\beta}=\lambda_{j,\beta;i,\alpha}$). The exchange symmetry of the equations of motion, inherited from the simulated bosonic system, ensures a symmetric state will always remain symmetric.

\begin{figure*}[!htbp]
    \centering
    \includegraphics[width=\linewidth,trim=0 10 0 10,clip]{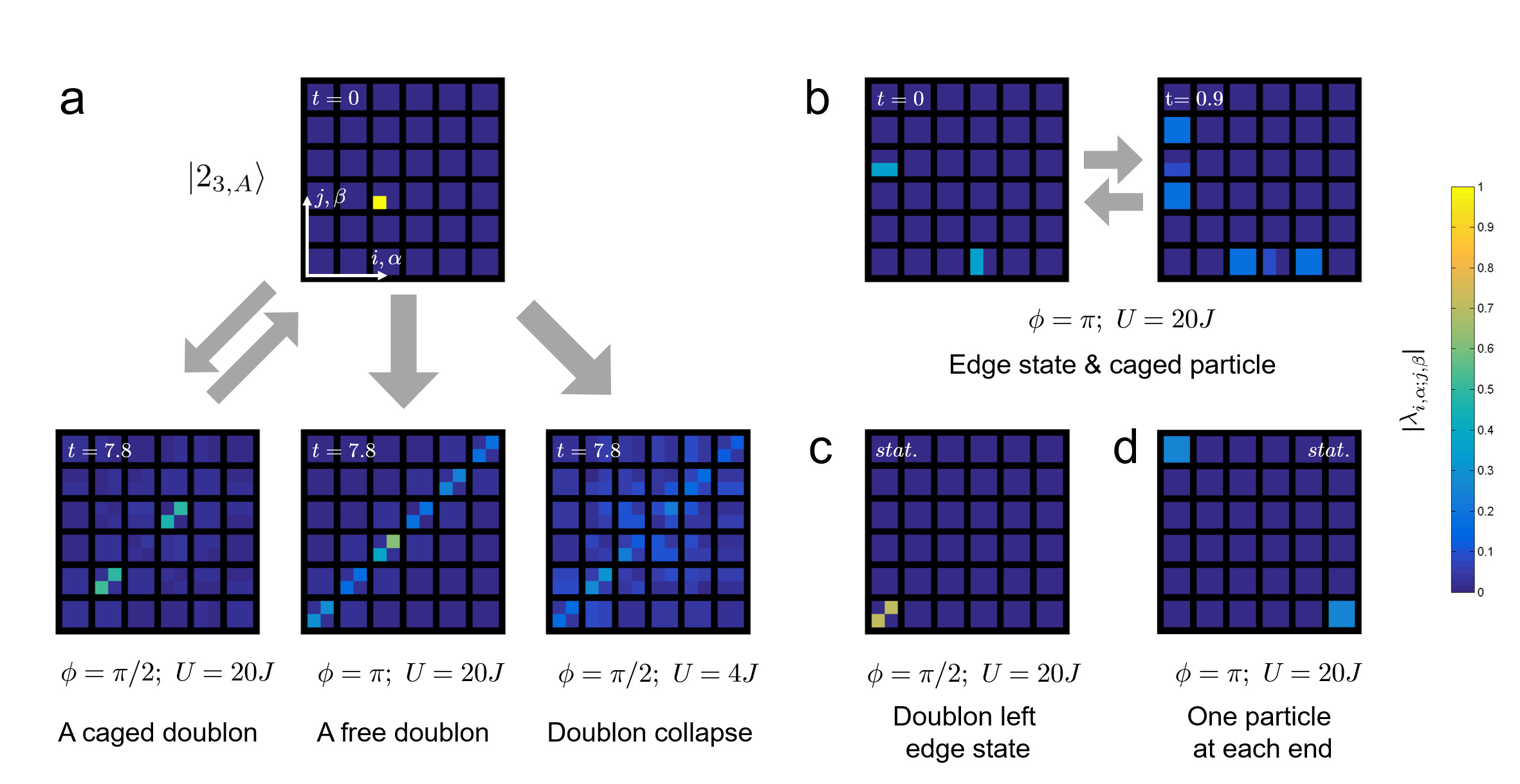}
    \caption{Quasi-2D plots for the simulations in Figure \ref{fig:5}. a) Different time evolutions for the initial state $\ket{2_{3,A}}$, depending on the parameters of the system (see Figure \ref{fig:5}b-d). b) Single particle edge state and caged particle (see Figure \ref{fig:5}e). c) Doublon edge state (Figure \ref{fig:5}f). d) One particle at each end (Figure \ref{fig:5}g). Note that colour corresponds to the first quantization components (see Equation \ref{eq:bases}), and hence the apparent preference for the diagonal in the doublon collapse plot, which actually corresponds to a more even distribution in the Fock basis (Figure \ref{fig:5}d.).}
    \label{fig:2D}
\end{figure*}

Although the resulting quasi-2D lattice might be thought of as four-dimensional, the small size of two of its dimensions can be exploited to rearrange the lattice in a 3D geometry with some nonlocal hopping amplitudes, as shown in Figure \ref{fig:6}b). To do this, it is useful to relabel the states with a new index $\zeta=1,2,3,4$, corresponding to the values $(\alpha,\beta)=(A,A),(A,B),(B,A),(B,B)$. This small perspective shift gives the key to a possible scheme for an experimental implementation of the model. In order to construct the equivalent of a 3D lattice evolving in time in a photonic system, at least one dimension must be synthetic. In the Creutz-Hubbard model, the natural choice is the $\zeta$ dimension described above.

This is especially convenient, because synthetic gauge fields and nonlocal hopping amplitudes are easier to implement in synthetic space than real space. \cite{Bell2017,Yuan2018} Three nonlocal hoppings would be necessary per unit cell (two for the rungless ladder), represented with dotted lines in Figure \ref{fig:6}b. It is worth noting that, if the synthetic dimension is periodic, no nonlocal hoppings are needed.

A specific setup for the photonic Creutz-Hubbard ladder could be similar to the one proposed in a recent work. \cite{Lustig2019} In it, oscillating columns of waveguides act as sites in a 2D lattice, and the synthetic dimension is represented by the modes of oscillation of the column. Four modes in a square lattice would suffice to implement our system.

In Figure \ref{fig:2D}, the quasi-2D representations of the simulations in Figure \ref{fig:5} are shown. This way of presenting them carries more information than its counterpart used in Figure \ref{fig:5}, but it is also much less intuitive. The lattice sites are arranged as indicated by Figure \ref{fig:6}a. In an experiment, the initial state $\ket{2_{i\alpha}}$ would be easy to prepare, allowing for the observation of the different phenomena in Figure \ref{fig:2D}a.

%SEC CONCLUSIONS
\section{Conclusions}
\label{sec:Con}

In this work, we have studied numerically several exotic phenomena that arise in the photonic Creutz and Creutz-Hubbard ladders. The former has already been implemented experimentally, and we propose a 3D waveguide lattice setup to implement the latter, possible thanks to the new possibilities that the use of synthetic dimensions can offer. It is our understanding that the current state of the art in the rapidly evolving field of photonics allows this proposal to be implemented in a real system. As far as we are aware, this would mark the first instance of a quasi-1D topological insulator with interactions in a waveguide-lattice-type quantum simulator, representing an increase in complexity compared to the strongly correlated systems that have been implemented so far in waveguide lattices (which were both 1D and topologically trivial).  \cite{Longhi2011,Krimer2011,Corrielli2013,Rai2015,Mukherjee2016}%<-Rev1 (Y PONERLES LO Q CAMBIAMOS PARA LA V2!!)

In the single-particle model, we have focused on the caging of particles and the topologically protected edge states, which are AB-localized in the absence of rung-like hoppings, even when no localized bulk energy eigenstates exist. We then considered the two-particle model, the minimal model with interactions, and studied the doublon collapse, the different caging regimes (in particular, the doublon caging for $m=0$, $\phi=\pi/2$) and the doublon edge states. These results show that the phase diagram of these systems have many interesting phenomena to observe away from the usually-studied $\phi=\pi$ limit. In particular, analogous caging regimes will be present for multiplons with different numbers of particles at different values of the magnetic flux, giving rise to complex physical situations that require further study.

Once realized, these photonic systems could be the first to observe many of the interesting features of the Creutz and Creutz-Hubbard models, marking a milestone in the research of low-dimensional topological systems and flat band models. Moreover, in the interacting case, it would mark a step forward in the understanding of the complicated interplay that occurs between interactions and topology, the details of which are still not completely understood. The Creutz and Creutz-Hubbard models constitute an ideal playground in which flat band dynamics, particle interactions and nontrivial topology can be studied in depth, separately or at the same time, allowing for a complete study of the rich interplay between them.\\ \\

\acknowledgments%Rev1
We would like to thank M. A. Mart\'in-Delgado, M. Bello and A. Berm\'udez for fruitful discussions. CEC was supported by Spain's MINECO through
Grant No. FIS2017-84368-P. GP was supported by Spain's MINECO through Grant No. MAT2017-
86717-P and by CSIC Research
Platform PTI-001.

%%%%%%%%%%%%%%%%%%%%%%%%%%%%%%%%%%%%%%%%%%%%%%%%%%%%%%%%
%APPENDICES
%%%%%%%%%%%%%%%%%%%%%%%%%%%%%%%%%%%%%%%%%%%%%%%%%%%%%%%%

\appendix

%APPENDIX A
\section{Numerical Zak phase calculation}
\label{app:Zak}

The Zak phase associated with a given energy band of a 1D system is always calculated in the bulk (i.e. with periodic boundary conditions), and is defined as:
\begin{equation}
    \sZ=i\oint_{BZ} \bra{k}\partial_k\ket{k} dk, \label{eq:Zak}
\end{equation}
where $\ket{k}$ is the state of the band with quasimomentum $k$ and the integral is calculated along the Brillouin zone.

To be able to do a numerical calculation of the Zak phase, we need to use the discretized version of expression (\ref{eq:Zak}):
\begin{equation}
    \sZ=i\sum_{k\in BZ} \frac{1}{\Delta k} \bra{k} \Big[\ket{k+\Delta k}-\ket{k}\Big] {\Delta k}, \label{eq:ZakD}
\end{equation}
where $\Delta k$ is the numerical reciprocal lattice spacing. For $\Delta k = \pi/La$, with $a$ the lattice constant and $L$ the length of the system, this is just the actual formula for which expression (\ref{eq:Zak}) is only a shorthand, given that the spatial periodicity of the bulk discretizes reciprocal space. Our reciprocal lattice spacing is $\Delta k = \pi/La$, and thus no approximation is made here. %Rev1: fixed lattice spacing and broken brakets

To do the numerical calculation, it is necessary to write equation (\ref{eq:ZakD}) in the form of a Wilson loop, because otherwise the result can be affected by the arbitrary gauge the numerically calculated states will inevitably have. Using the Taylor series for the logarithm we obtain:
\begin{equation}
    \sZ = -\mathrm{Im} \log \prod_{k\in BZ} \braket{{k}}{{k+\Delta k}}, \label{eq:ZakComp}
\end{equation}

where we take the principal value of the logarithm. This causes the Zak phase to be defined modulo $2\pi$. Equation (\ref{eq:ZakComp}) was used in the numerical calculation mentioned in the main text.

%APPENDIX B
\section{Equations of motion}
\label{app:Syst}

Using the first-quantized basis for the states and the definitions in Equations (\ref{eq:CreutzH}) and (\ref{eq:CHubbH}) for the Hamiltonian, the Schr\"odinger equation for the two-particle Creutz-Hubbard system, $\sH_{C\! H}\ket{\psi}=E\ket{\psi}$, turns into the following system of linear equations: %<-Rev1 First quantized
\begin{align}
    &E \lambda_{i,\alpha;j,\beta} = -J[\tilde{\lambda}_{i+1,\bar\alpha;j,\beta}+\tilde{\lambda}_{i-1,\bar\alpha;j,\beta}+\tilde{\lambda}_{i,\alpha;j+1,\bar\beta}\nonumber\\
    &+\tilde{\lambda}_{i,\alpha;j-1,\bar\beta}]-J_\alpha\tilde{\lambda}_{i-s_\alpha,\alpha;j,\beta}-J^*_\alpha\tilde{\lambda}_{i+s_\alpha,\alpha;j,\beta}\nonumber\\
    &-J_\beta\tilde{\lambda}_{i,\alpha;j-s_\beta,\beta}-J^*_\beta\tilde{\lambda}_{i,\alpha;j+s_\beta,\beta}\nonumber\\
    &-m[\tilde{\lambda}_{i,\bar\alpha;j,\beta}+\tilde{\lambda}_{i,\alpha;j,\bar\beta}]   \kern 40 pt       \textrm{for }(i,\alpha)\ne(j,\beta)\nonumber\\    
    &\frac{E}{2\sqrt{2}} \lambda_{i,\alpha;i,\alpha} = -J[\tilde{\lambda}_{i+1,\bar\alpha;i,\alpha}+\tilde{\lambda}_{i-1,\bar\alpha;i,\alpha}]\nonumber\\
    &-J_\alpha\tilde{\lambda}_{i-s_\alpha,\alpha;i,\alpha}-J^*_\alpha\tilde{\lambda}_{i+s_\alpha,\alpha;i,\alpha}\nonumber\\
    &-m\tilde{\lambda}_{i,\bar\alpha;i,\alpha}         +\frac{U}{2}\tilde{\lambda}_{i,\alpha;i,\alpha},   \label{eq:syst}
\end{align}
with $\tilde{\lambda}_{i,\alpha;j,\beta}\equiv[1+ \delta_{i,j}\delta_{\alpha,\beta}(\sqrt{2}-1)]\lambda_{i,\alpha;j,\beta}$.

As discussed in the main text, this system of equations can be reinterpreted as describing the movement of one particle on a quasi-2D lattice with the coordinates $(i,\alpha,j,\beta)$.

%%%%%%%%%%%%%%%%%%%%%%%%%%%%%%%%%%%%%%%%%%%%%%%%%%%%%%%%
%BIBLIOGRAPHY
%%%%%%%%%%%%%%%%%%%%%%%%%%%%%%%%%%%%%%%%%%%%%%%%%%%%%%%%
\bibliographystyle{aipnum4-1}
\bibliography{Refs_brief}

\end{document}